\documentclass[review,1p,times]{elsarticle}   	
\usepackage[utf8]{inputenc}
\usepackage[english]{babel}
\usepackage{amsmath}
\usepackage{amsfonts}
\usepackage{amssymb}
\usepackage{graphicx}
\usepackage{bm}
\usepackage{dcolumn}

\usepackage{lineno}
\usepackage[unicode]{hyperref}
\usepackage{latexsym}
\usepackage{geometry} 
\usepackage{url}
\usepackage{color}

\usepackage{enumitem}

\usepackage{booktabs}
\journal{Physic Letters B}
\begin{document}
\begin{frontmatter}
\title{Measurement of the \texorpdfstring{$^{154}$Gd}{Gd}(n,$\gamma$) cross section and its astrophysical implications}
\author[1,2]{A. Mazzone}
\author[3,4]{S. Cristallo}
\author[5]{O. Aberle}
\author[6]{G. Alaerts}
\author[7]{V. Alcayne}
\author[8,9]{S. Amaducci}
\author[10]{J. Andrzejewski}
\author[11]{L. Audouin}
\author[12]{V. Babiano-Suarez}
\author[5,13,14]{M. Bacak}
% ...
\cortext[cor1]{Corresponding Author}

\address[1]{Istituto Nazionale di Fisica Nucleare, Bari, Italy }
\address[2]{Consiglio Nazionale delle Ricerche, Bari, Italy}
\address[3]{Istituto Nazionale di Fisica Nazionale, Perugia, Italy }
\address[4]{Istituto Nazionale di Astrofisica - Osservatorio Astronomico d'Abruzzo, Italy }
\address[5]{European Organisation for Nuclear Research (CERN), Switzerland }
\address[6]{European Commission, Joint Research Centre, Geel, Retieseweg 111, B-2440 Geel,Belgium }
\address[7]{Centro de Investigaciones Energ\'eticas Medioambientales y Tecnol\'ogicas (CIEMAT),Spain }
\address[8]{INFN Laboratori Nazionali del Sud, Catania, Italy }
\address[9]{Dipartimento di Fisica e Astronomia, Universit\`a\ di Catania, Italy }
\address[10]{University of Lodz, Poland }
\address[11]{IPN, CNRS-IN2P3, Univ. Paris-Sud, Universit\'e\ Paris-Saclay, F-91406 Orsay Cedex,France }
\address[12]{Instituto de F\'\i sica Corpuscular, CSIC - Universidad de Valencia, Spain }
\address[13]{Technische Universit\"at Wien, Austria }
\address[14]{CEA Saclay, Irfu, Universit\'e\ Paris-Saclay, Gif-sur-Yvette, France }
% ...
\begin{abstract}
The neutron capture cross section of $^{154}$Gd was measured from 1 eV to 300 keV in the experimental area located 185 m from the CERN n\_TOF neutron spallation source, using a metallic sample of gadolinium, enriched to 67$\%$ in $^{154}$Gd. The capture measurement, performed with four C$_{6}$D$_{6}$ scintillation detectors, has been complemented by a transmission measurement performed at the GELINA time-of-flight facility (JRC-Geel), thus minimising the uncertainty related to sample composition.
An accurate Maxwellian averaged capture cross section (MACS) was deduced over the temperature range of interest for s process nucleosynthesis modeling. We report a value of 880(50) mb for the MACS at $kT=30$ keV, significantly lower compared to values available in literature. 
The new adopted $^{154}$Gd(n,$\gamma$) cross section reduces the discrepancy between observed and calculated solar s-only isotopic abundances predicted by s-process nucleosynthesis models.
\end{abstract}

\begin{keyword}
s process \sep $^{154}$Gd \sep Neutron time of flight \sep n\_TOF
\end{keyword}
\end{frontmatter}
\section{Introduction}
All the elements heavier than those in the iron group are produced by a sequence of neutron capture reactions and $\beta$ decays taking place in a hot stellar environment, during different phases of stellar evolution. The two main processes involved are the slow (s) and the rapid (r) neutron capture processes. The s process \cite{B2FH,Cameron_1957} owes its name to the neutron-capture time scale, which allows $\beta$ decay to occur between consecutive capture events. Consequently, a series of these reactions produce stable isotopes by moving along the $\beta$-stability valley. On the other hand, when the neutron densities are high enough \cite{thielemann2011astrophysical}, the neutron capture sequence is much faster than the $\beta$ decays and the path, the r process path, can proceed toward many short-lived isotopes, approaching the neutron drip line.

\begin{figure}[htb]
\center{\includegraphics[width=220 pt]{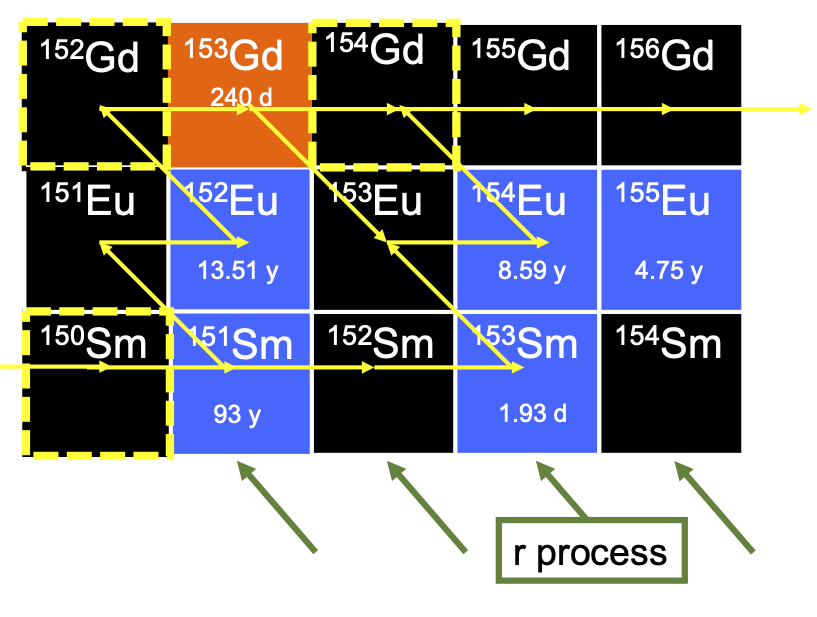}}
\caption{(Color online) The yellow line represents the main s-process path in the Sm-Eu-Gd region; the s-only isotopes are highlighted with a yellow dashed box. Stable elements are in black, $\beta^+$, $\beta^-$ and $\beta^+$ radioisotopes are in orange, blue and red, respectively.}
\label{fig:s-path}
\end{figure}

Most nuclei receive a contribution from both the s and the r processes (see e.g. \cite{prantzos2019chemical}). However, a few isotopes cannot receive any contribution from the r process because they are shielded against $\beta$ decays by stable isobars and for this reason are called s-only isotopes. 
This is the case of the two gadolinium isotopes $^{152}$Gd and $^{154}$Gd which are shielded against the $\beta$-decay chains from the r-process region by their stable samarium isobars, as shown in Figure \ref{fig:s-path}. To be precise, $^{152}$Gd may receive an additional contribution from the p process, which proceeds via photo-disintegration. The amount of the p-process contribution to the $^{152}$Gd abundance is still far from being precisely determined (see \cite{travaglio2018role} and references therein), while a minor p-process contribution to the $^{154}$Gd abundance cannot be excluded as well.

The almost pure s-process origin of $^{154}$Gd (as for other s-only isotopes), makes its capture cross section crucial for testing various stellar models aiming at understanding s process nucleosynthesis in Asymptotic Giant Branch (AGB) stars, the most important stellar site for the s process. In particular, relevant hints on the shape and extension of the main s process neutron source, the so-called $^{13}$C pocket \cite{Gallino_1998}, can be derived. Recently, several studies \cite{bisterzo2014galactic,cristallo2015need,prantzos2019chemical} have been conducted analyzing the solar s process abundances in the framework of a Galactic Chemical Evolution (GCE) model to investigate the effect of different internal structures of the $^{13}$C pocket, which may affect the efficiency of the $^{13}$C($\alpha$,n)$^{16}$O reaction. In addition, Trippella and Collaborators~\cite{trippella2014s} carried out a similar analysis based on single stellar models. Cristallo et al.~\cite{cristallo2015need} and Prantzos et al.~\cite{prantzos2019chemical} have obtained an under-production (-30$\%$) of the $^{154}$Gd\footnote{In the following, an individual isotope or/and its abundance is imply indicated by the symbol (e.g. $^{152}$Gd), to simplify the notation.} s-only nucleus compared to the s-only isotope $^{150}$Sm, which is an un-branched isotope, usually assumed as a reference for the s process flow. This result is at odds with observations. The authors of ref. \cite{cristallo2015need}, have suggested that part of such a deviation could be connected to uncertainties in the adopted nuclear physics inputs, taken from the Karlsruhe Astrophysical Database of Nucleosynthesis in Stars (KADoNiS) version 0.3~\cite{KADONIS}, including the neutron capture cross section of $^{154}$Gd. In the work by Trippella et al.~\cite{trippella2014s}, problems for the s process production of Gd were found as well, although in this case, the discrepancy between the abundances of $^{150}$Sm and $^{154}$Gd is less pronounced. The need for a new $^{154}$Gd(n,$\gamma$) measurement was underlined by the unreasonable prediction for the over-productions of $^{152}$Gd and $^{154}$Gd with respect to their solar abundances. In particular, the ratio $^{154}$Gd/$^{152}$Gd), turned out to be lower than unity, while it is thought that $^{152}$Gd should exhibit a higher p-process contribution as compared to $^{154}$Gd \cite{trippella2014s}. In addition, $^{154}$Gd was found to be produced insufficiently compared to the lighter s-only $^{148}$Sm and $^{142}$Nd, produced mostly by the s process and possibly partly by the p process.

The close correlation between stellar abundances and neutron capture cross sections calls for an accurate determination of the $^{154}$Gd(n,$\gamma$) cross section. In addition, the reduction of the uncertainty related to nuclear physics inputs could rule out one of the possible causes of present discrepancies between observation and model predictions of the abundances. In fact, refined stellar models require a full set of Maxwellian Averaged Capture Cross Section (MACS) for thermal energies in the range $kT= 8 - 30$ keV. In the case of $^{154}$Gd, 80\% of the MACS at $kT = 8$ keV is determined by the capture cross section in the neutron energy region between $2.7 - 300$ keV - hereafter we refer to this energy region as Unresolved Resonance Region (URR). At $kT = 30$ keV, the MACS depends almost entirely on the cross section in the URR. In this energy region, three time-of-flight $^{154}$Gd(n,$\gamma$) cross section measurements are reported in literature, by Shorin {\it et al.} \cite{shorin1974neutron}, Beer and Macklin \cite{beer1988sm} and Wisshak {\it et al.} \cite{wisshak1995stellar}. They all cover the URR and the respective MACS exhibit large differences and at $kT= 30$ keV, they range from 878(27) mb to 1278(102) mb. Therefore, the data present in the literature, so far, are not conclusive enough to constrain stellar model calculations. 

The large spread in the available experimental data could be related to the corrections for isotopic impurities which are necessarily applied in the data analysis. In particular, the poor knowledge of their cross sections, given the low natural abundance of $^{154}$Gd (2.18$\%$). Different discrepancies may be related to: $\textit{i)}$ the detectors used in the past experiments, which in some cases could have suffered from high neutron sensitivity; $\textit{ii)}$ the experimental determination of the neutron flux, which might have been biased in some previous measurements; $\textit{iii)}$ the quality of oxide samples and the need of canning for the container to avoid loss of material.

The present measurement reduced the impact of these limiting factors, by using the well-established, low neutron-sensitivity C$_{6}$D$_{6}$ detectors \cite{mastinu2013new}, combined to a self-sustaining metallic sample enriched in $^{154}$Gd, and exploiting the results of the recent $^{155}$Gd(n,$\gamma$) measurement performed at n$\_$TOF \cite{mastromarco2019cross}. Moreover, the gadolinium sample was characterised by a transmission measurement at the neutron time-of-flight facility GELINA at EC-JRC-Geel (Belgium).
\section{Measurements}
The neutron capture cross section measurement was performed at the neutron time-of-flight facility n$\_$TOF at CERN. In this facility, neutrons are produced by spallation reactions induced on a lead target by 20 GeV/c protons from the CERN Proton Synchrotron (PS), which provides a total of $2 \times 10^{15} $ neutrons/pulse, generated by a 7$ \times 10^{12}$ protons/pulse primary beam. The initially fast neutrons are moderated and then collimated through two flight paths of different lengths. The present measurement was performed at the experimental area located 185 m from the spallation target. This long flight-path, combined with the 7 ns width of the proton bunches from the PS, results in a high energy resolution ranging from 3$\times 10^{-4}$ at 1 eV to 3$\times 10^{-3}$ a 100 keV \cite{guerrero2013performance}.
The neutron capture events were observed via the detection of the prompt $\gamma$-ray cascade from $^{155}$Gd excited states. Four C$_{6}$D$_{6}$ detectors, modified for minimizing their neutron sensitivity \cite{mastinu2013new}, were arranged at 125$^{\circ}$ relative to the neutron beam direction and about 10 cm upstream from the gadolinium sample position. This configuration minimised the effect of anisotropic emission of $\gamma$ cascades while reducing the background from in-beam photons scattered by the sample. The Total Energy detectors (see \cite{SCHILLEBEECKX20123054} and references therein) were used in combination with Pulse Height Weighting Technique (PHWT) \cite{borella2007use,SCHILLEBEECKX20123054}. 

The sample consisted of 0.263 g of metallic gadolinium enriched at 66.78$\%$ in $^{154}$Gd. The main contaminant, i.e. $^{155}$Gd, was declared by the sample provider at 17.52$\%$. A detailed resonance analysis of the capture data, based on the recent results obtained by time of flight on $^{155}$Gd (n,$\gamma$) at n$\_$TOF, allowed us to estimate a content of $^{155}$Gd equal to 20.2$\%$. This higher value was confirmed by a transmission measurement on the same sample carried out at GELINA. 

A Au sample of the same diameter was used to normalise the measured yield, by applying the saturated resonance technique \cite{macklin1967capture}. Also, two other samples with the same diameter of 3 cm were used. A lead sample enabled the estimate of the background, while a natural gadolinium sample allowed to assign observed resonances to the correct gadolinium isotope, besides confirming the isotopic content of $^{154}$Gd in the enriched sample.

As mentioned above, the gadolinium sample was further studied through the transmission measurement at a 10-m station of the GELINA facility. The transmission, $T$, which was experimentally obtained from the ratio of Li-glass spectra resulting from a sample-in and a sample-out measurement, is related to the total cross section $\sigma_{tot}$ by the equation:
\begin{equation}
\label{eq::NDS}
T(E_n)=e^{-n\sigma_{tot}(E_n)},
\end{equation}
where $n=(1.431\pm0.006)\times10^{-4}$ atoms/b denotes the areal density of the gadolinium sample.
GELINA is particularly suitable for high-resolution transmission measurement, because of its time characteristic and the small dimensions of the neutron producing target. For this experiment, the neutron beam was collimated to a diameter of 10 mm at the sample position and filters were placed near the sample to absorb slow neutrons from the previous neutron-burst and to continuously monitor the background.  The neutron beam passing through the sample was detected by a 6.4 mm thick and 76 mm wide Li-glass scintillator enriched to 95\% in $^{6}$Li. The detector was placed at 10.86 m from the neutron production target.

\section{Data analysis and results}
The experimental capture yield \textit{Y$_{c}$}, i.e. the probability for an incident neutron to be captured in the sample, can be deduced from the measured count rate, corrected for the detection efficiency of capture events. 
By applying the PHWT, the count rate, C$_w$, is weighted in order to make the detection efficiency independent of the cascade path and $\gamma$ multiplicity. The weighting functions were calculated simulating the response of the full apparatus by a GEANT4 \cite{agostinelli2003geant4} Monte-Carlo simulation. The capture yield can be written as:
\begin{equation}
\label{eq::yield}
Y(E_{n})=N\frac{C_{w}(E_{n})-B_{w}(E_{n})}{\Phi(E_{n})}
\end{equation}
where \textit{N} is a normalisation factor, \textit{B$_{w}$} is the weighted count rate representing the background and \textit{$\Phi$} is the neutron flux impinging on the sample.
The neutron energy \textit{E$_{n}$} was determined from the measured time of flight using an effective flight path determined from well-known low energy resonances in Au \cite{massimi2011neutron}.

The normalisation factor groups together geometrical factors, such as the area of the sample and its beam-interception factor, the solid angle subtended by the capture and flux monitors, and the detection efficiency. It was obtained with the saturated resonance technique applied to the 4.9-eV resonance in $^{197}$Au. This  normalisation, based on the Au capture data, was within 1.3\% consistent with the normalisation derived from a fit to the capture data using the parameters from the transmission data.

The background, \textit{B$_{w}$}, includes different contributions: \textit{i)} ambient background, which was determined from a measurement in absence of the neutron beam; 
\textit{ii)} the sample-independent background (also referred to as empty background), due to the neutron beam, which was estimated from a measurement with neutron beam impinging on an empty sample;
\textit{iii)} the sample-dependent background, either due to sample-scattered neutrons (subsequently captured in the environmental material and generating $\gamma$-rays in the experimental area) or due to $\gamma$-rays produced at the spallation target and reaching the experimental area, where they can be scattered by the sample into the detectors. This third component was estimated by a measurement with a lead sample in the neutron beam. Previous measurements and Monte-Carlo simulations showed that the contribution of the in-beam $\gamma$-ray background is relevant only in a limited energy window of 1-100 keV~\cite{guerrero2013performance}. Therefore, it was possible to disentangle the two sample-dependent background components in the time-of-flight spectrum measured with the lead and with the gadolinium samples, properly scaled. The neutron background was scaled for the elastic cross section and the areal density of the gadolinium and lead sample, while the $\gamma$ background was scaled for the effective atomic number of gadolinium and lead samples. The same procedure for the estimation of the total background was adopted in the study of the $^{197}$Au(n,$\gamma$) cross section. In particular, in the energy region from 5 keV to 500 keV, where this cross section is very well-known (and above 200 keV, considered as a standard) the capture cross section from this study resulted to be in good agreement with evaluations.
 
\begin{figure}[htb]
\center{\includegraphics[width=220 pt]{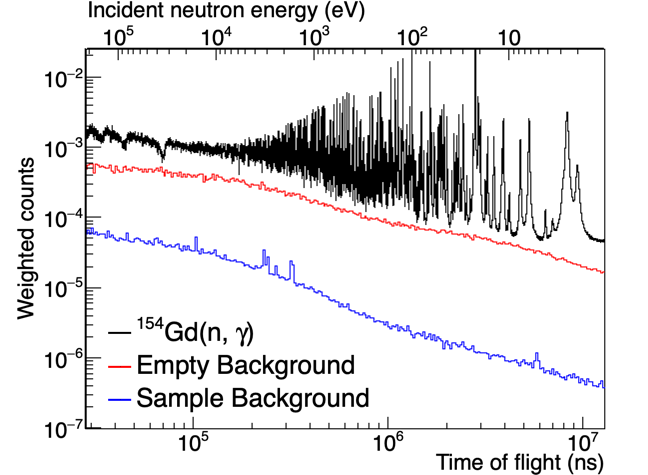}}
\caption{ (Color online) Weighted spectrum measured with the $^{154}$Gd sample compared to the background measured with an empty-sample holder and the one estimated from a measurement with a lead sample. The last spectrum was obtained by scaling the in-beam $\gamma$-ray background and the component due to the neutrons scattered by the sample, see text for details. }
\label{fig:BKG}
\end{figure}

In figure \ref{fig:BKG} the weighted number of counts, registered with the $^{154}$Gd sample, are shown together with the sample-independent (empty) background and the sample-dependent background components. The empty-sample background component dominates the total background over the energy region of interest, whereas the sample-dependent background contributes by less than 4$\%$ to the total background. The absolute magnitude of the background was additionally verified with dedicated runs using black resonance filters~\cite{SCHILLEBEECKX20123054} 
and a fair agreement was found. In the region between 5 and 300 keV, the signal-to-total background ratio is 2.5. Although the total background level is significant, its uncertainty is small since the background is dominated by the empty-sample component, which is known within 1$\%$.

The neutron flux was evaluated with a dedicated measurement campaign using different detection systems and with neutron cross section standards \cite{barbagallo2013high}. 
The estimated systematic uncertainty on the flux determination was within 1$\%$ below 3 keV \cite{barbagallo2013high} and of 3.5 $\%$ up to 300 keV.

\begin{figure}[htb]
\center{\includegraphics[width=220 pt]{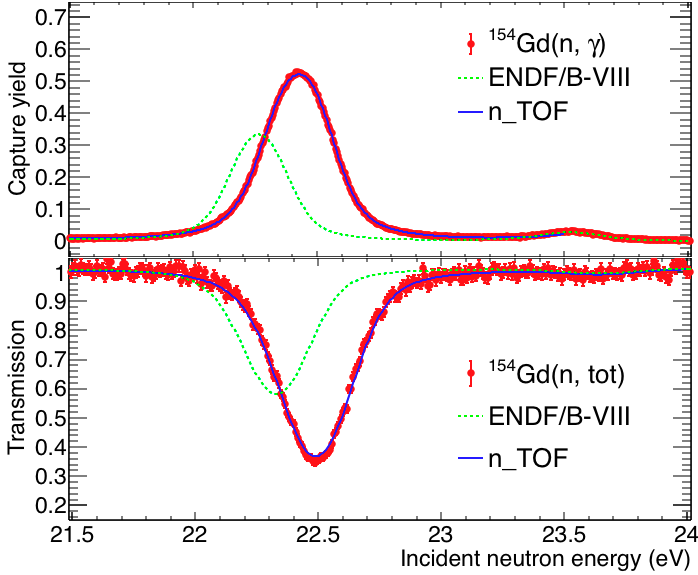}}
\caption{(Color online) Measured capture yield and transmission of for n + $^{154}$Gd. Experimental data are shown by red symbols, the n$\_$TOF R-matrix fit in blue and the yield calculated from ENDF/B-VIII.0 parameters in green, both in continuous lines.}
\label{fig:RRR}
\end{figure}

In the energy region up to 2.7 keV -  hereafter referred to as Resolved Resonance Region (RRR) -, the experimental capture yield was analysed with the Bayesian R-matrix analysis code SAMMY \cite{larson1998updated}. The code can manage experimental effects such as Doppler and resolution broadening, self-shielding and multiple interactions of neutrons in the sample. 
Sizable discrepancies were found for some neutron resonances compared to the yields obtained using the ENDF/B-VIII.0 evaluation data set \cite{brown2018endf}. These differences were further confirmed by the transmission data. An example is shown in figure \ref{fig:RRR}, where the present results of resonance shape analysis are compared to the expected values obtained using resonance parameters from the evaluated data set.  
Up to 2.75 keV, we analysed 156 resonances and 3 new resonances were found at 183.17(2), 197.65(1) and 376.26(1) eV, respectively. The statistical analysis of resonance parameters (similar to ref.~\cite{mastromarco2019cross}) yielded a neutron strength function $S_0  = 2.78(33) \times 10^{-4}$, an average level spacing $D_0= 13.4(11)$ eV and an average radiation width $\overline{\Gamma_\gamma}=59(7)$ meV.

Above this energy region, the experimental resolution became too low to resolve individual resonances and an averaged cross section was determined in the neutron energy range from 2.7 keV to 300 keV. The capture yield was corrected for multiple scattering and self-shielding through Monte-Carlo simulations as described in ref.~\cite{mingrone}. The resulting correction for the two effects was lower than 2$\%$. 

\begin{figure}[htb]
\center{\includegraphics[width=220 pt]{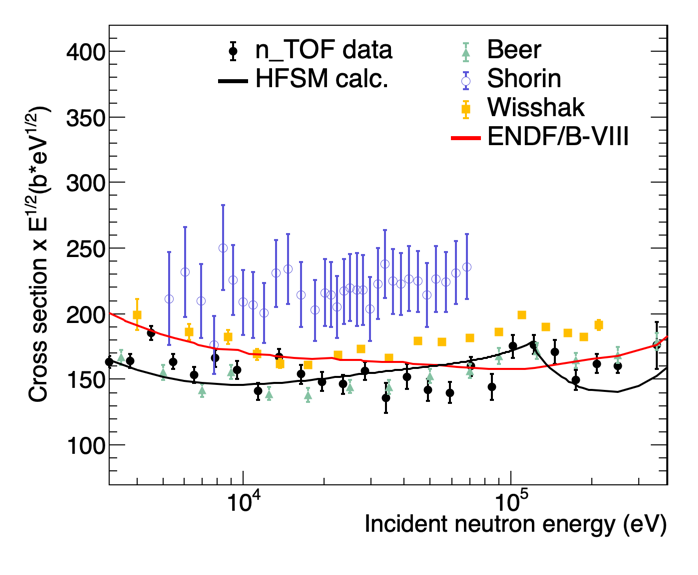}}
\caption{(Color online) $^{154}$Gd(n,$\gamma$) cross section from the present study compared to previous measurements (coloured symbols) and evaluation (continuous line). The HF calculation has been normalized by a factor 0.737 (see text).}
\label{fig:URR}
\end{figure}

The contribution of the contaminants present in the measured samples was taken into account in the experimental data analysis. In particular, for the main contaminant $^{155}$Gd, the cross section was assumed from the previous n$\_$TOF measurement on $^{155}$Gd. In figure \ref{fig:URR}, the capture cross section extracted from this study is compared to the data of Shorin et al.~\cite{shorin1974neutron}, Beer and Macklin~\cite{beer1988sm}, Wisshak et al.~\cite{wisshak1995stellar} and the ENDF/B-VIII evaluation. In the URR, the present data fairly agree with the data by Beer and Macklin \cite{beer1988sm} and they are substantially lower than the data reported by Wisshak et al. \cite{wisshak1995stellar} and Shorin et al. \cite{shorin1974neutron}. This comparison seems to indicate that the results obtained with similar experimental setups are in fair agreement, while they are inconsistent if the adopted measurement technique is different. A discussion of potential reasons for the disagreement is beyond the scope of this article.

MACS as a function of the thermal energy kT were calculated from the present capture data in the RRR and in the URR. The cross section in the energy region above this range (i.e. above 300 keV) was taken into account by calculations performed using the Hauser Feshbach (HF) statistical model theory, as implemented in the TALYS code~\cite{TALYS}. The average resonance parameters, obtained in the present analysis of the RRR, were constrained to be reproduced by the calculations by adjusting the level density and the $\gamma$-ray strength function. An overall normalisation by a factor 0.737 of the resulting capture cross section was still necessary to reproduce the present experimental MACS at $kT = 30$ keV.

The uncertainty on the MACS takes into account the uncorrelated uncertainty attributable to counting statistics and systematic uncertainties. The uncertainty components originate from the normalisation of the capture data and the PHWT (1.3$\%$), the shape of the neutron flux (1$\%$ below 3 keV and 3.5$\%$ above) and the subtraction of the background (on average 2$\%$, depending on the energy region.). Another minor uncertainty is associated with the alignment of the sample and its geometrical shape. As a result, over the thermal energy range of kT$=5-100$ keV, the uncertainty on the MACS ranges between 5 and 7$\%$.
\begin{table}[ht!]
        \caption{Maxwellian Averaged Capture Cross (in mb) calculated for the n$\_$TOF data in the energy range kT$=5-100$ keV. The values are compared with those reported in the literature by KADoNiS 0.3 and KADoNiS 1.0. }
        \label{tab:MACS}
        \centering
		\begin{tabular}{c|ccc}
        \toprule
         Energy & n$\_$TOF & KADoNiS & KADoNiS \\
         (keV)& & 0.3&1.0\\
        \midrule
        5 & 2160(90) & 2801 & 2947(190)\\
        \midrule
        8 & 1820(80) & -- & 2195(87)\\
        \midrule
        10 & 1590(80) & 1863 & 1924(61)\\
        \midrule
        15 & 1270(80) & 1477 & 1537(33)\\
        \midrule
        20 & 1080(60) & 1258 & 1326(24)\\
        \midrule
        25 & 970(50) & 1124 & 1188(19)\\
        \midrule
        30 & 880(50) & 1028(12) & 1088(16)\\
        \midrule
        40 &  770(40) & 898 & 950(14)\\
       \midrule
        50 & 690(40) & 810 & 856(12)\\
        \midrule
        60 & 640(40) & 745 & 786(12)\\
         \midrule
        80 & 560(30) & 653 & 690(15)\\
         \midrule
        100 & 510(30) &591 & 626(19)\\
        \bottomrule
        \end{tabular}
\end{table}
In table \ref{tab:MACS}, the present MACS are reported for the thermal energy grid proposed by KADoNiS \cite{KADONIS,dillmann2010kadonis}. In the whole energy range, the MACS values from our measurement are significantly lower than KADoNiS. It is interesting to note that the disagreement worsened with the new version of the evaluation, the deviation being between 10$\%$ and 20$\%$.
\section{Astrophysical implications}
As discussed above, a new determination of the $^{154}$Gd neutron capture cross section was motivated by a discrepancy between stellar models and observations, as highlighted by \cite{cristallo2015need} and confirmed by \cite{prantzos2019chemical}, where a lower theoretical $^{154}$Gd/$^{150}$Sm ratio with respect to that measured in the Sun (and derived for the early-solar system) was found. In fact,
theoretical values, which include yields from the Asymptotic Giant Branch (AGB) phase of low and intermediate mass stars (taken from the FRUITY database of AGB star nucleosynthesis \cite{cristallo2011evolution,cristallo2015evolution}), show an under-production of $^{154}$Gd with respect to $^{150}$Sm: $^{154}$Gd/$^{150}$Sm = 0.70. As a consequence, one can argue that a reason for the disagreement should be attributed to problems in the cross section of $^{154}$Gd itself. 

The use of the present cross section leads to an increase of the theoretical solar $^{154}$Gd abundance by 10\% on average\footnote{Note that only a limited number of AGB models have been investigated with the present cross section. The evaluation of the effect in a full GCE model will be published separately.}. 
The difference in the $^{154}$Gd surface abundances is lower than the change of the neutron capture cross sections (on average 15\% compared to KaDoNiS 0.3). This is because the $^{154}$Gd production/destruction strongly depends on the branching at $^{154}$Eu, which is an unstable isotope (its decay lifetime in the terrestrial condition is 8.6 y). This branching is by-passed when the major neutron source in AGB stars (the $^{13}$C($\alpha$,n)$^{16}$O reaction) is activated, due to its short lifetime for the timescale characterizing neutron captures in this regime. The situation may be different during thermal pulses when the higher temperature can efficiently activate the $^{22}$Ne($\alpha$,n)$^{25}$Mg neutron source (which produces a definitely higher neutron flux). In such a case, the neutron capture channel is competitive compared to the $\beta$ decay channel and, as a consequence, the main s-process path may by-pass $^{154}$Gd. The delicate balance between neutron captures on $^{154}$Gd and $\beta$ decays from $^{154}$Eu determines the final abundance of $^{154}$Gd.

In summary, the present experimental value leads to a better agreement between model calculations and observations, although it is not able to completely remove the mismatch. In 2009, Lodders and collaborators~\cite{lodders2009landolt} estimated the uncertainty in the determination of the solar gadolinium\footnote{No info on isotopic uncertainties are currently available.} abundance to be $\pm$15\% (and $\pm$5\% for samarium). The adoption of the present $^{154}$Gd neutron capture cross section, eventually leads to a new $^{154}$Gd/$^{150}$Sm ratio of 0.77 in FRUITY models. When taking into consideration the lower limit of the present neutron capture cross section, we obtain $^{154}$Gd/ $^{150}$Sm $\simeq$ 0.81. As a consequence, the ratios obtained in FRUITY models are now compatible with the observed abundances, within observational uncertainties ($\pm$20\%). 

When the present cross section is used in the models by Trippella et al.~\cite{trippella2014s,trippella2016s}, it is interesting to notice that while the discrepancy with respect to $^{142}$Nd, $^{148}$Sm and $^{152}$Gd is completely erased (due to the larger production of $^{154}$Gd), at the same time, the ratio $^{154}$Gd/$^{150}$Sm attains a value (1.15) consistent with observations, within uncertainties. 
In general, it appears that the approach in ref. \cite{trippella2016s} produces a flatter distribution of s process isotopes, although this was obtained in post-process computations and not in full stellar models. Therefore, a clear suggestion emerging from the present $^{154}$Gd(n,$\gamma$) cross section measurement is that some of the remaining model ambiguities might be solved by a merging of the mixing approaches presented in FRUITY and ref.~\cite{trippella2014s}, something that is in an advanced stage of implementation \cite{Vescovi}.

Another important outcome from this combined experimental and theoretical study is related to the
$^{154}$Gd abundance, which largely depends on the branching at $^{154}$Eu. For this isotope, no experimental data are available for both the neutron capture cross section and the temperature-dependent $\beta$ decay rate. Therefore, s process calculations are based on purely theoretical estimations (see ref.~\cite{rauscher2000astrophysical} and ref.~\cite{takahashi1987beta}, respectively). A lower $^{154}$Eu(n,$\gamma$) cross section or a faster $^{154}$Eu($\beta^-$)$^{154}$Gd decay would lead to a larger $^{154}$Gd surface abundance with respect to $^{150}$Sm (and vice-versa). Therefore, the present result suggests that additional efforts should be spent in this direction from the experimental side, to provide experimental values as detailed as possible to stellar modelers.

\section{Acknowledgements}
The isotope used in this research was supplied by the United States Department of Energy Office of Science by the Isotope Program in the Office of Nuclear Physics.

This research was supported by the EUFRAT open access programme of the Joint Research Centre at Geel.

\bibliography{article_Gd154}

\end{document}